\documentclass[preprint,12pt]{elsarticle}

\usepackage{graphicx}
\usepackage{amssymb}
\usepackage{amsmath}
\usepackage{subcaption}

\usepackage{lineno}

\usepackage[utf8]{inputenc}

\usepackage{array}
\newcolumntype{P}[1]{>{\centering\arraybackslash}p{#1}}

\let\vec\mathbf

\journal{Journal of Electrostatics}

\begin{document}

\begin{frontmatter}


\title{Atomistic Field Theory for Contact Electrification of Dielectrics}


\author[2]{Khalid M. Abdelaziz}
\address[2]{Department of Mechanical and Nuclear Engineering, Kansas State University}
\author[1]{James Chen\corref{label2}}
\cortext[label2]{Corresponding Author; chenjm@buffalo.edu}
\address[1]{Department of Mechanical and Aerospace Engineering, University at Buffalo -- The State University of New York}
\author[2]{Tyler J. Hieber}
\author[2]{Zayd C. Leseman}

\begin{abstract}
The triboelectrification of conducting materials can be explained by electron transfer between different Fermi levels. However,  triboelectrification in dielectrics is poorly understood. The surface dipole formations are shown to be caused by the contact-induced surface lattice deformations. An Atomistic Field Theory (AFT) based formulation is utilized to calculate the distribution of the polarization, electric and potential fields. The induced fields are considered as the driving force for charge transfer. The simulation results show that a MgO/BaTiO$_3$ tribopair can generate up to 104 $V/cm^2$, which is comprable with the data in the published literature.
\end{abstract}


\end{frontmatter}


\section{Introduction}
\label{S:1}
The triboelectric effect or contact electrification is an experimentally proven phenomenon \cite{Sow2012}. Its occurrence in conducting materials can be explained by electron transfer resulting from the difference in work functions or Fermi levels of the contacting metals. That is, electrons in a metal with a higher energy level lower their energy by moving to a metal with a lower energy level \cite{Vasandani2017, Hogue2004}. However, when a dielectric material is involved, the essential cause of the charge transfer is largely debatable \cite{Sow2012}: is it that rubbing the two surfaces increases the microscopic area of contact, or that it contributes energy to affect the charge transfer \cite{Sow2012}. Additionally, the mechanism of the charge transfer is also debatable: is it the migration of electrons \cite{Liu2008,Liu2009} or ions \cite{McCarty2007, Diaz1993} or material "pieces" from one surface to another \cite{Sow2013}. Because the fundamental cause and mechanism are not known, the answers to this very question about the exhibited behaviors of tribopairs involving dielectrics remain unclear: For example, there is no definite explanation for which dielectric material will attain a positive or a negative charge when it comes in contact with another. Furthermore, even for a given pair of materials, the direction of charge transfer cannot be reliably predicted \cite{Wang2017b}. To answer these questions, different empirical Triboelectric Series  \cite{Diaz2004}, which present an ordering of the materials depending on their tendency to attain positive or negative charges upon contact, have been developed. However, the actual exhibited behavior can depend on a multitude of factors that are not taken into account when the series are developed, which makes them unreliable \cite{Lowell1980}. In fact, experiments have shown that factors including the nature of contact \cite{Baytekin2012}, temperature \cite{Lu2017}, surface defects \cite{Mukherjee2016}, the presence of adsorbates in the air \cite{Byun2016} and the material strain \cite{Wang2017b} greatly affect the results of triboelectrification experiments.

The occurrence of charge transfer necessitates the occurrence of a difference between the potentials of the surfaces in contact. Assuming defect-free surface lattices, unstrained materials and that the experiment is performed in vacuum; prior to any material, ionic or electronic migrations, the only remaining factor that can affect the surface potentials upon contact would be the formation of surface dipoles \cite{Smith1969, Mnch2001}. Therefore, this work postulates that the cause of triboelectrification or contact eelctrification in dielectrics is attributed to the contact-induced surface lattice deformations which result in the formation of surface dipoles. Furthermore, an Atomistic Field Theory (AFT) based \cite{Chen2005} formulation is presented to efficiently calculate the distribution of the polarization, the electric potential and field and the charge density given the state of the constituent atoms of the surface lattices. MD simulations are used to simulate the lattice deformations resulting from the contact of Perovskite crystalline structure Barium Titanate (BaTiO$_3$) and Magnesia (MgO) because these materials have well established models in the literature \cite{Vielma2013,Chen2011}. It is shown that lattice deformations occur when the two materials are placed in sufficient proximity for the atomic interactions across the boundary to become strong enough to alter the atomic positions and form the surface dipoles.

Although the detailed mechanism of triboelectrification is still poorly understood, it has been the core of several different applications. Triboelectric Nanogenerators (TENGs) are an application of the triboelectric effect that has recently been drawing a lot of attention \cite{Jung2015,Sim2016,Zhu2016,Dhakar2016}. A TENG is able to convert mechanical to electrical energy similar to other energy harvesting devices but has a high volume energy density (490 $kW/m^3$ \cite{Niu2015}) which makes it an attractive alternative for utilizing wasted mechanical energy. A TENG utilizes dielectric materials, such as Perovskite-structure BaTiO$_3$ \cite{Wang2017} and Polytetrafluoroethylene (PTFE) \cite{Wang2017,Yang2015}, as triboelectric pairs which underlines the need to further understand triboelectricity in dielectrics. Being dielectric, the materials trap the induced charge rather than transfer it. Consequently, the trapped charge creates an electric field, which induces electrical charge transfer in neighboring electrodes made of conducting materials \cite{Niu2015}. Because of the coarse-grain nature of AFT, the presented formulation has potential to model the actual size of a TENG device (at the $\mu$m scale) and strengthens its suitability as a design tool for TENGs and other triboelectric devices.

Section 2 of this work derives the developed atomistic formulation and the approach to obtain the electric characteristics from the simulation results. Section 3 illustrates the atomistic models of the materials utilized in the MD simulation. Section 4 describes, in detail, the simulation procedure. Section 5 discusses the obtained dipole formations and electric characteristics. The conclusions can be found in Section 6.

\section{Atomistic Formulation for Electromechanical Coupling}
\label{S:2}

In an MD simulation, atomic forces are calculated at each time step and the positions are updated by time integration. Using these positions, the electrical characteristics (electric field, electric potential and charge density) can be calculated by iterating relevant formulas \cite{Griffiths2005} over all the atoms in the system. To simulate a micro-scale triboelectric layer that is usually employed in a TENG, a relatively large number of atoms is needed that will make the calculation be extremely time consuming.

Chen et. al. introduced a concept of dipole formation for lattices at their current state and hypothesized that each lattice can be represented as a dipole \cite{Chen2010b,Chen2011b}. In other words, a crystalline structure is approximated as a collection of dipole. This dipole can not only be used for the calculations of the electric characteristics but also related to the Miller indices for crystallines. A perfect lattice in general results to no polarization relative to the center of the lattice. However, if a lattice is perturbed by stimuli, e.g. temperature, mechanical forces or body forces, the motions of all atoms within a lattice consequently produce polarization and the lattice can be considered as a dipole. It should be noticed that this concept is different from a molecular dipole. The dipole for a lattice is defined at its perturbed state while a molecular dipole  exists in the ground state of a molecule. 

Each consequent dipole represents a lattice and induces an electric field in its proximity. The surface lattices of both materials are interfered by each other when the pair is in close distance during electrification. In this work, the induced field is hypothesized as the driving force for charge transfer. 

There are multiple theories to describe the lattice properties. Atomistic Field Theory (AFT) \cite{Chen2005} is one approach to efficiently obtain properties from the state of the atoms at a certain time. Solids possess a repetitive pattern of atoms referred to as the Bravais lattice, which is neutrally charged. By placing a node at the center of a representative lattice, the motion of any atom within the lattice can be expressed by \cite{Chen2010b,Chen2011b}:
\begin{equation}
\label{eq:aftmotion}
\vec u(k,\alpha) = \vec u(k)+\vec{\zeta}(k,\alpha)
\end{equation}
where $\alpha$ and $k$ represent the $\alpha$-th atom in the $k$-th unit cell, $\vec{u}(k)$ is the displacement of the $k$-th unit cell and $\vec{\zeta}(k,\alpha)$ is the relative displacement of atom $\alpha$ to the centroid of the $k$-th unit cell. 
All the physical quantities can be then expressed in physical and phase spaces, which are connected through the Dirac delta function, $\delta$, and the Kronecker delta function, $\tilde{\delta}$, as
\begin{equation}
A(\mathbf{x},\mathbf{y}^\alpha, t)=\sum_{k=1,}^{N_{uc}}\sum_{\alpha=1}^{N_a} a[\mathbf{r}(t), \mathbf{p}(t)]\delta(\vec{R}^k-\vec x)\tilde{\delta}(\Delta r^{k\zeta}-y^{\alpha})
\end{equation}
with normalization conditions
\begin{equation}
\int_{V^*}\delta(\vec{R}^k-\vec x)d^3\mathbf{x}=1\qquad(k=1,2,3,...,n)
\end{equation}
where $V^*$ is the volume of a unit cell; $\vec{R}^k$ and $\vec x$ is the position vector of the $k$-th unit cell in the phase and physical spaces, respectively. $N_{uc}$ and $N_a$ are the number of unit cells in the system and the number of atoms in the k-th unit cell, respectively. 

It is straightforward to define polarization density, $\mathbf{p}(\mathbf{x},\mathbf{y}^\alpha, t)$, of $\zeta$-th atom within $k$-th unit cell as
\begin{equation}
\mathbf{P}(\mathbf{x},\mathbf{y}^\alpha, t)=\sum_{k=1,}^{N_{uc}}\sum_{\alpha=1}^{N_a} q^\zeta(\mathbf{R}^k+\Delta r^{k\zeta}) \delta(\vec{R}^k-\vec x)\tilde{\delta}(\Delta r^{k\zeta}-y^{\alpha})
\end{equation}
By averaging over the unit cells results in the homogeneous field, the polarization density, $\mathbf{P}(\mathbf{x}, t)$, for the unit cell at the position $\vec{x}$ is given by \cite{Chen2010b,Chen2011b}:
\begin{equation}
\label{eq:sumpol}
\vec P(\vec x, t) = \sum_{k=1,}^{N_{uc}}\sum_{\alpha=1}^{N_a} q^{\alpha} \vec{d}^{k\alpha}\delta(\vec{R}^k-\vec x)
\end{equation}
where $q^{\alpha}$ is the charge of atom $\alpha$ and $\vec{d}^{k\alpha}$ is the displacement (relative to the center of the lattice) of the $\alpha$-th atom in the $k$-th unit cell. 

When the lattices of the two materials approach each other, the constituent atoms of both lattices interact (repulse/attract) according to the assumed interatomic potential. 
Such atomistic motions result in dipole formation on the surface. The electric potential density at the position $\vec z$ due to the unit cell at $\vec x$ could be calculated from \cite{Chen2010b,Chen2011b,Wang2013}:
\begin{equation}
\label{eq:dipolepotential}
V(\vec z, \vec x, t)=\sum_{k=1}^{N_{uc}}\sum_{\alpha=1}^{N_a} q^{\alpha} \vec{d}^{k\alpha}\cdot\frac{(\vec z-\vec x)}{{|\vec z-\vec x|}^3}\delta(\vec{R}^k-\vec x)
\end{equation}
$N_{uc}$ is the number of unit cells in the system instead of the atoms, which considerably improves the calculation performance. Consequently, the induced electric field density at the position $\vec z$ by a unit cell located at $\vec x$ can be calculated from by using $E=-\nabla_\vec{z} V$ \cite{Chen2010b,Chen2011b,Wang2013}:
\begin{equation}
\label{eq:fieldfrompotential}
\vec E(\vec z, \vec x, t)=\sum_{k=1}^{N_{uc}}\sum_{\alpha=1}^{N_a} q^{\alpha} \vec{d}^{k\alpha}\cdot \left(\frac{3(\vec z-\vec x)\otimes(\vec z-\vec x)}{{|\vec z-\vec x|}^5}-\frac{\vec I}{{|\vec z-\vec x|}^3}\right)\delta(\vec{R}^k-\vec x)
\end{equation}
where $\vec I$ is the identity matrix. The electric field at position $\mathbf{z}$ induced by all unit cells can be found by integrating Equation~\ref{eq:fieldfrompotential} over all unit cells as \cite{Chen2010b,Chen2011b}
\begin{align}
\vec E(\vec z, t)=\int\sum_{k=1}^{N_{uc}}\sum_{\alpha=1}^{N_a} & q^{\alpha} \vec{d}^{k\alpha}\cdot \nonumber\\
 & \left(\frac{3(\vec z-\vec x)\otimes(\vec z-\vec x)}{{|\vec z-\vec x|}^5}-\frac{\vec I}{{|\vec z-\vec x|}^3}\right)\delta(\vec{R}^k-\vec x)d^3\mathbf{x}
\end{align}

\section{Material Choice}
\label{S:3}

The test case involves a Perovskite crystalline structure barium titanate (BaTiO$_3$), and a rocksalt crystalline structure magnesia (MgO). Both are modeled using the Coulomb-Buckingham potential \cite{Vielma2013, Shukla2008, Chen2010,Buckingham1938}:
\begin{equation}
\label{eq:coulbuck}
U^{ij}(r^{ij}) = \frac{q^iq^j}{r^{ij}}+Ae^{\frac {-r^{ij}} {\rho}}-\frac{C}{{r^{ij}}^6}
\end{equation}
where $U^{ij}$ is the potential, $r^{ij}$ is the interatomic distance, $q^i$ is the charge of the $i$-th atom and $A$ and $\rho$ and $C$ are species-to-species dependent parameters \cite{Vielma2013}.

\citet{Chen2011} showed that for the original Coulomb-Buckingham potential shown in Equation \ref{eq:coulbuck}, an unphysical collision between oxygen atoms (Buckingham Catastrophe) can occur when the interatomic distance becomes lower than a critical value. Therefore, the modification suggested by \citet{Chen2011} is included: The addition of a Lennard-Jones ${r^{ij}}^{-12}$ repulsive term. The final form of the potential becomes:
\begin{equation}
\label{eq:coulbuckcat}
U^{ij}(r) = \frac{q^iq^j}{r^{ij}}+Ae^{\frac {-r^{ij}} {\rho}}-\frac{C}{{r^{ij}}^6}+\frac{D}{{r^{ij}}^{12}} \quad\quad r^{ij} < r_c
\end{equation}
where $D$ can assume the same value of $C$ \cite{Chen2011}. In all cases, the value of the potential $U^{ij}$ is assumed to equal 0 when $r^{ij}$ is greater than a pre-specified cutoff $r_c$. As a result, the interatomic force is given by:
\begin{equation}
\label{eq:coulbuckforce}
{\vec {F}}^{ij} = -\frac{\partial U}{\partial r^{ij}}=r^{ij}\lbrack\frac{q^iq^j}{{r^{ij}}^3} + \frac{1}{\rho r^{ij}} A^{ij} e^{-\frac{{r}^{ij}}{\rho}} - 6C^{ij}r^{-8} + 12D^{ij}r^{-14}\rbrack \quad\quad r^{ij} < r_c
\end{equation}
Since the force formulation in Equation \ref{eq:coulbuckforce} now combines both the mechanical and electrical effects, the equation of motion can now be written as:
\begin{equation}
\label{eq:atommotion}
m_i\ddot{\vec x}_i(t) = \sum_{j=1,j\neq i}^{N_a} {\vec {F}}^{ij}
\end{equation}
where $N_a$ represents all the atoms within the cutoff distance $r_c$. Table \ref{table:species} lists the atomic properties utilized in the test case while Table \ref{table:potential} lists the modified Coulomb-Buckingham potential parameters for the involved species, where pairs like Mg-Ti are assumed to have only a Coulombic interaction\cite{Vielma2013,Chen2011}.

\begin{table}[h]
\centering
\begin{tabular}{P{2.5cm} P{2.5cm} P{2.5cm}}
    \hline
    \textbf{Species}  & \textbf{Mass (u)}  & \textbf{Charge (e$^-$)} \\ \hline
    Ba  & 137.327 & 2 \\
    Ti  & 47.867  & 4 \\
    O   & 15.999  & -2 \\
    Mg  & 24.305  & 2 \\
  \end{tabular}
\caption{Mass and charge values for the involved atom types}\label{table:species}
\end{table}

\begin{table}[h]
\centering
\begin{tabular}{P{2.5cm} P{2.5cm} P{2.5cm} P{3cm}}
    \hline
    \textbf{Pair}  & \textbf{A (e$^-$V)}  & \textbf{B (\AA)} & \textbf{C (e$^-$V $\AA^{-6}$)} \\ \hline
    Ba-O  & 1588.36 & 0.3553  & 0 \\
    Ti-O  & 3131.25 & 0.2591  & 0 \\
    O-O   & 2641.4 & 0.3507  & 535.37 \\
    Mg-O  & 8216.6  & 0.3242  & 0 \\
  \end{tabular} 
\caption{Coulomb-Buckingham potential parameter values for the involved materials \cite{Vielma2013,Chen2011}}\label{table:potential}
\end{table}

\section{Simulation Procedure}
\label{S:5}

This section illustrates the simulation case from which various results are drawn in the following section. The case uses a quasi-static simulation approach to exclude any transient effects from the results. The case is run using the LAMMPS \cite{Plimpton1995} MD simulation package and visualized using OVITO \cite{Stukowski2009}. To calculate the dipole moment vector $\vec P$, the atoms are grouped into identical groups which are initially neutrally charged because of their symmetry. When the atomic positions shift from the neutral position during the simulation, $\vec P$ for each group is calculated as described in Section \ref{S:2}.

\begin{figure}[h!]
\centering\includegraphics[width=0.8\linewidth]{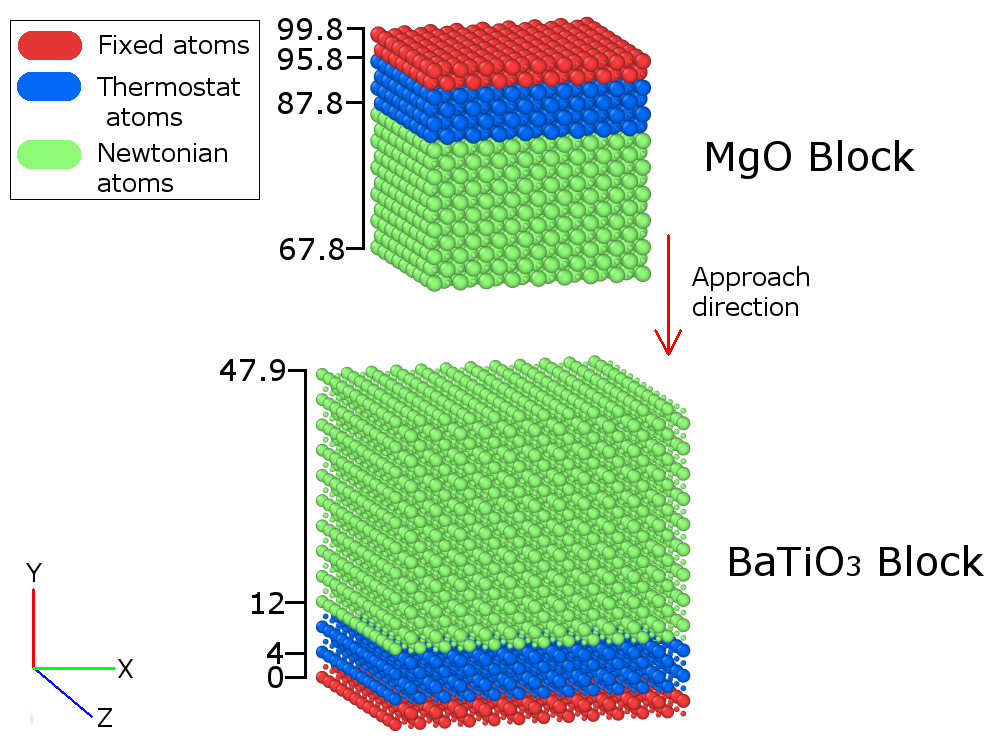}
\caption{Initial setup of the MgO slab (upper) and the BaTiO$_3$ slab (lower). Dimensions are in Angstroms along the y-direction}\label{fig:case1}
\end{figure}

Figure \ref{fig:case1} shows the initial setup of the simulation case. The MgO (upper) and BaTiO$_3$ (lower) slabs are initially positioned at a distance from each other to equilibrate independently. The separation distance is set higher than the interatomic potential cutoff distance to guarantee this effect. The BaTiO$_3$ slab, as well as the simulation domain, have periodic boundary conditions in the $x$ and $z$ directions and a fixed boundary in the $y$ direction to simulate the approach. The MgO slab is finite-size to simulate a smaller object electrifying a larger material slab and to be able to generate a variation of the properties on the surface of the BaTiO$_3$ slab. The thermostat layers in both slabs are used to control the temperature of the system by means of a Nosé-Hoover thermal bath \cite{Melchionna1993}.

In the beginning, the atoms of both slabs are positioned in the simulation box \cite{Chen2010b} at a temperature of 0 $K$ with a separation distance of 5 lattice constants. With a simulation time step of 1 fs, the temperature is allowed to rise to room temperature (300 $K$) in 20,000 time steps. Additionally, the temperature is held at 300 $K$ for another 20,000 time steps to remove any effects of the temperature rise and achieve equilibrium. The number of equilibration time steps is always determined by allowing the maximum interatomic force to become nearly constant.

As previously mentioned, the rest of the simulation follows a quasi-static scheme. After equilibration, the MgO slab is shifted to a proximity of 2 lattice constants from the BaTiO$_3$ slab. This is followed by 5,000 time steps of equilibration. Thereafter, the MgO slab is shifted 0.2 lattice distance towards the BaTiO$_3$ slab followed again by 1,000 time steps of equilibration. The process is repeated until the nominal separation distance between both slabs vanishes, i.e. total displacement equals 2 lattice constants after thermal equilibration. However, an actual separation still exists due to the repulsion between the atoms (see Figure \ref{fig:distortion}). This final approaching step is also followed by 20,000 time steps of equilibration. Similar to the thermal equilibration, the number of time steps for equilibration is determined by allowing the maximum interatomic force to reach a constant.

\section{Results and Discussion}

\subsection{Dipole Formation} 

\begin{figure}[h!]
\centering\includegraphics[width=0.7\linewidth]{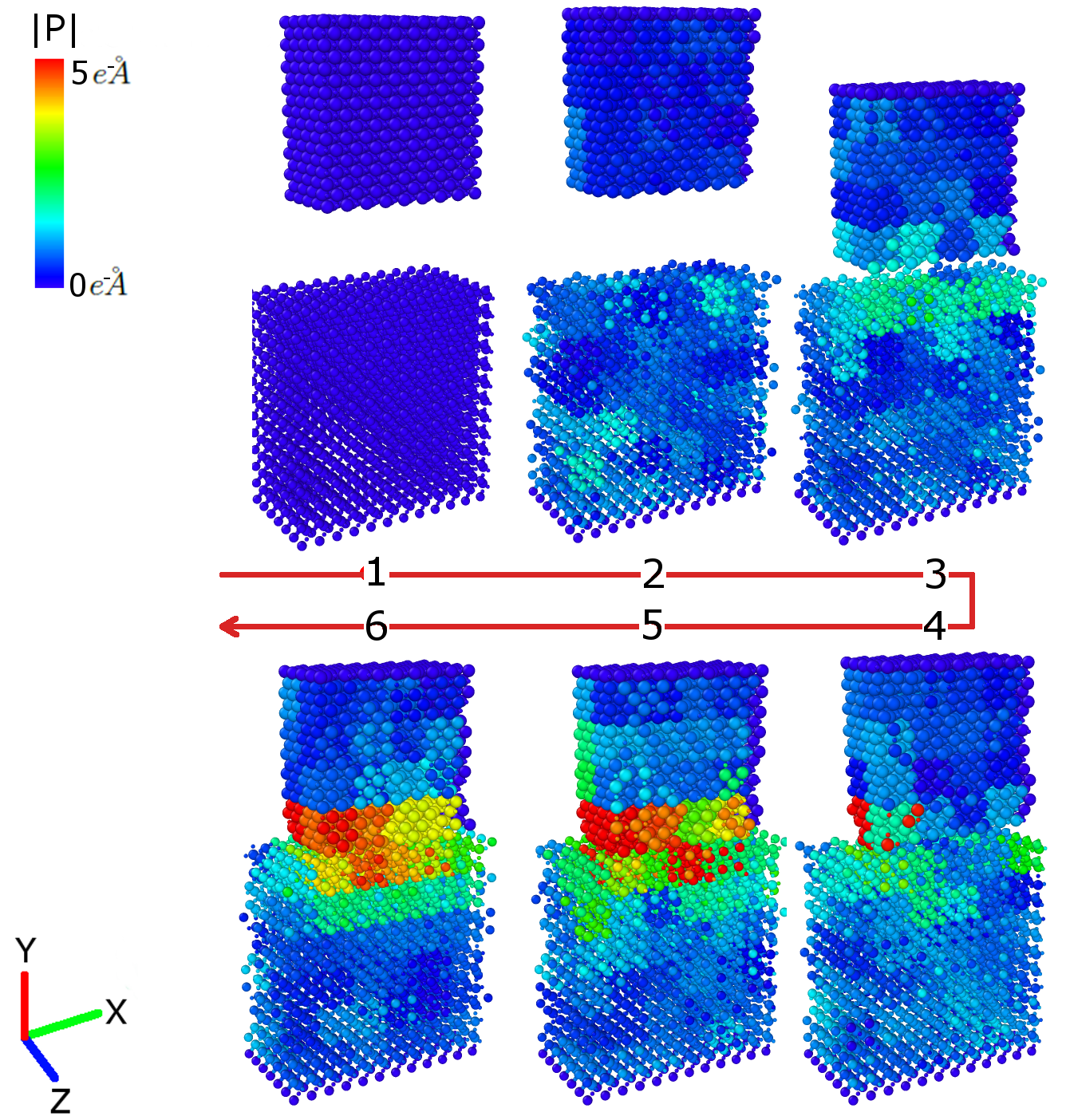}
\caption{Evolution of the dipole magnitude (Equation \ref{eq:sumpol}): (1) initial state, (2) after equilibration, (3) approach, (4) tilt of MgO slab due to attraction, (5) smallest gap, (6) after final equilibration}\label{fig:dipoleevolution}
\end{figure}

As the magnesia moves toward the barium titanate, dipoles form on the surface lattice  under the influence of atomic interaction. Figure \ref{fig:dipoleevolution} shows the evolution of the dipole magnitude value using a number of key frames of the simulation. It is noted that the BaTiO$_3$ atoms have a tendency to form dipoles during equilibration (large atomic oscillations) unlike MgO which is relatively stable. Also; in frames 4,5 and 6; the lattices on the surface of both slabs have a considerably higher dipole magnitude which underlines the dominance of the surface rather than the bulk effect in triboelectricity. It can be seen in frame 6 that the formed surface dipoles persist after equilibration which ensures they are not caused by transient effects.

\begin{figure}[h!]
	\begin{subfigure}[t]{0.45\textwidth}
    \centering
		\includegraphics[width=0.7\textwidth]{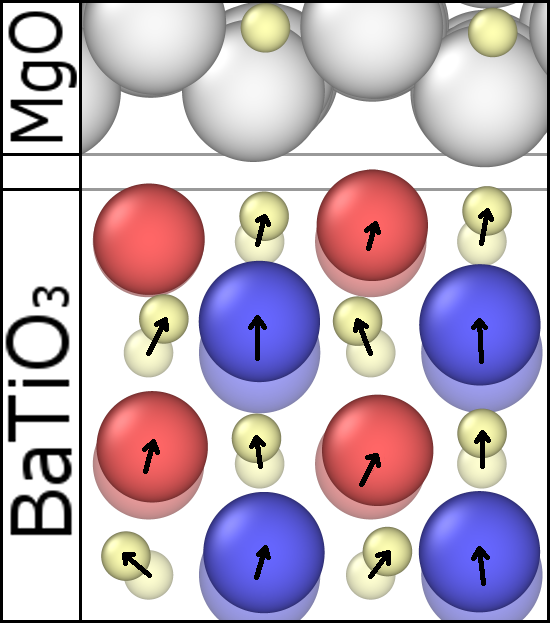}
        \caption{ }
		\label{fig:distortion}
	\end{subfigure}
	\begin{subfigure}[t]{0.45\textwidth}
        \centering
		\includegraphics[width=0.7\textwidth]{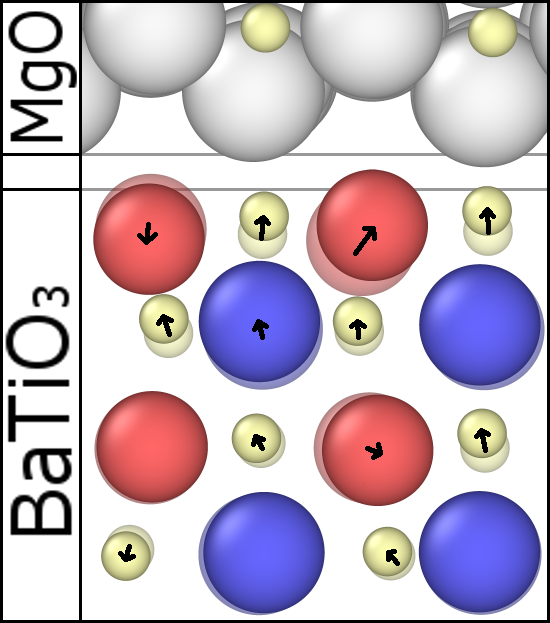}	
        \caption{ }
		\label{fig:distortionrelative}
    \end{subfigure}
    \caption{BaTiO$_3$ lattice deformation due to the proximity of the MgO atoms after 50,000 time steps (a) relative to neutral position (b) relative to the previous frame after 48,000 time steps. The arrows represent the atomic displacements. Yellow sphere are O$^{2-}$, white spheres are Mg$^{2+}$, red spheres are Ba$^{2+}$ and blue spheres are Ti$^{4+}$.}
\end{figure}

Figure \ref{fig:distortion} isolates a group of surface atoms from both material slabs to illustrate the change in the atomic positions relative to their original (neutral) positions. Figure \ref{fig:distortionrelative} shows the same comparison between the frame at 48,000 time steps and the frame at 50,000 time steps, which attains a higher dipole magnitude value. It is when the Mg-O distance across the two materials becomes smaller than the Mg-O distance within the MgO slab that the formation of the dipole is most pronounced. At this point, the attraction between Mg (charge $2 e^-$) and O from BaTiO$_3$ (charge $-2 e^-$) causes a considerable distortion of both lattices. These findings establish the connection between the lattice distortions and the formation of the surface dipoles prior to any material/ion/electron migration between the blocks.

\subsection{Electric Characteristics}

\begin{figure}[h!]
\centering\includegraphics[width=0.6\linewidth]{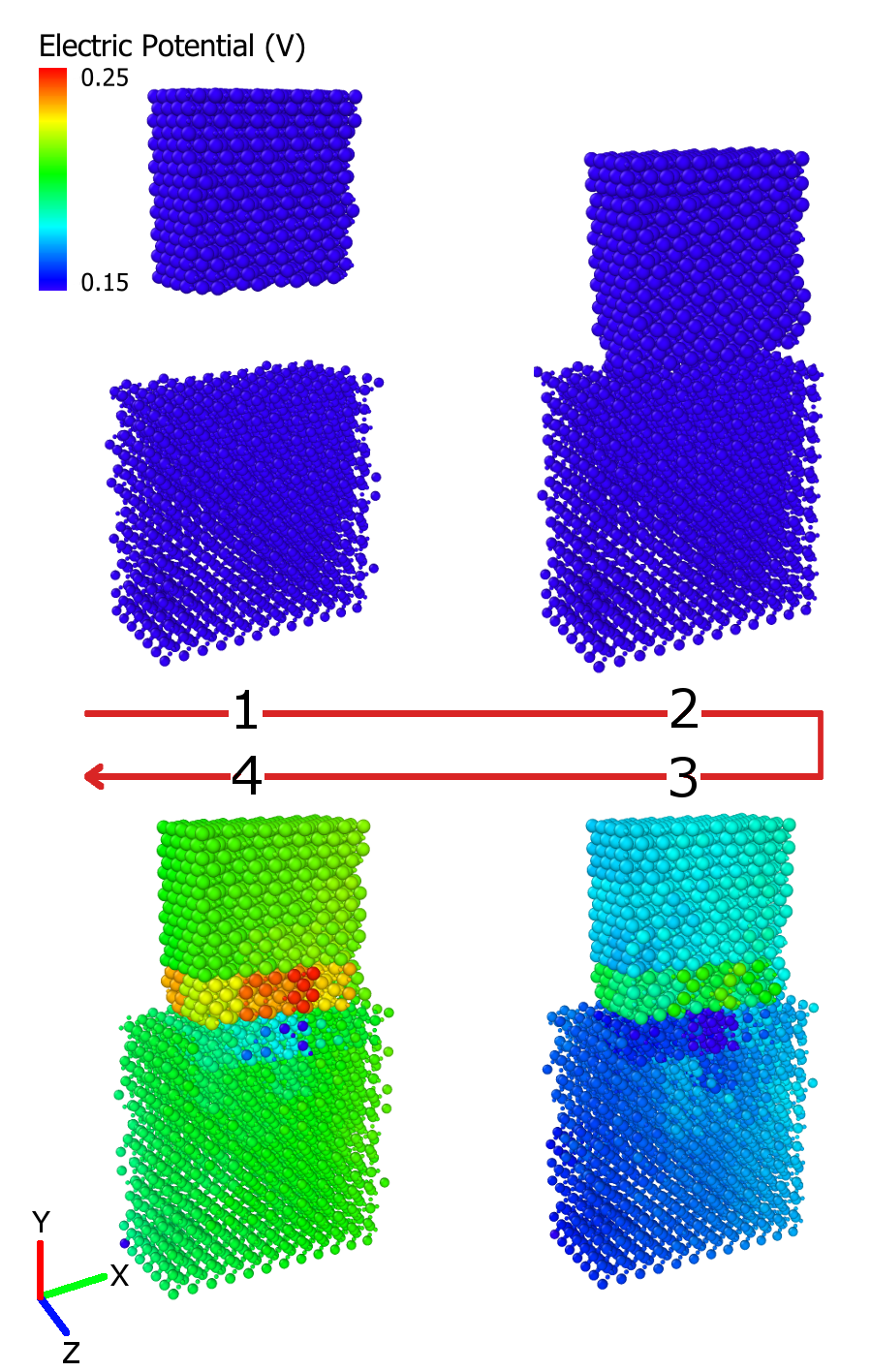}
\caption{Evolution of the electric potential: (1) after equilibration, (2) tilt of MgO block due to attraction, (3) smallest gap, (4) after final equilibration}\label{fig:potentialevolution}
\end{figure}

Each dipole from the unit cell induces a field in the neighborhood of the dipole. By summing over all dipoles from representative unit cells for a given point, the local electric potential (Equation \ref{eq:dipolepotential}), at such point can be calculated. Figure \ref{fig:potentialevolution} shows the electric potential distribution for some of the key frames discussed in Figure \ref{fig:dipoleevolution}. The formation of the dipoles due to lattice distortions illustrated in the previous section is confirmed to result in an alteration of the surface potentials of both material slabs which is a necessary precondition to charge transfer and contact electrification. This comes in support for the postulation that surface dipole formations contribute to contact electrification or triboelectrification in dielectrics. Due to the oscillations of the BaTiO$_3$ atoms, the magnitudes and polarities of the induced potential differ between the frames, but a potential difference between either sides and across the ends of the blocks always exists. To isolate the effect of these oscillations, the evolution of the potential difference between points of interest in both slabs was studied. During equilibration, a negligible potential difference is attained due to the random motions of the atoms. In the approach stage, the average values of the potential difference increases. After the approach is complete, the average values increase further up to 5 mV. Assuming a modest linear correlation between the size of the blocks and the observed potential difference between the ends, a density of the potential difference is found to be 104 $V/cm^2$. This behavior confirms that the effect of BaTiO$_3$ atom oscillations on the surface potentials is negligible when compared with the contact-induced surface lattice deformations and the accompanying dipole formations. 
A recent experiment on hybrid piezo-triboelectric generator with polytetrafluroethelene (PTFE) and organic ferroelectric polyvinylidene (PVDF). The triboelectric pair of PVDF and gold has shown the voltage output as high as 370 $V/cm^2$ and is capable of powering 600 LED bulbs \cite{Jung2015}. The potential difference density found in this study compares well with the experimental data for ferroelectric materials, i.e. BaTiO$_3$ and PVDF. It also confirms that triboelectrification and contact electrification produces higher output voltage than piezoelectric effect and others.
However, the discrepancy between the two values could be the result of the utilization of different materials and composite material structures in addition to the absence of the air gap resistance from the MD model. Another possibility attributed to the discrepency between experiments and simulation is the crystal size. Infinite 2D plane, i.e. periodic in x and z directions, is assumed in this study.

\begin{figure}[h!]
\centering\includegraphics[width=\linewidth]{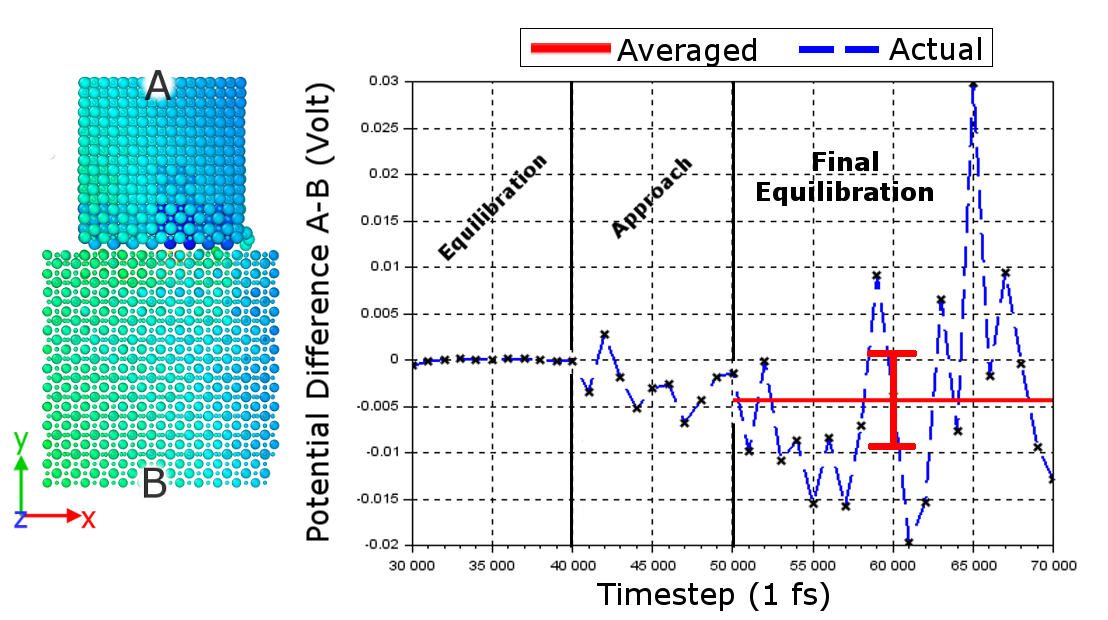}
\caption{Evolution of the electric potential difference along the Y direction}\label{fig:potentialdiffevolutiony}
\end{figure}

\begin{figure}[h!]
\centering\includegraphics[width=0.8\linewidth]{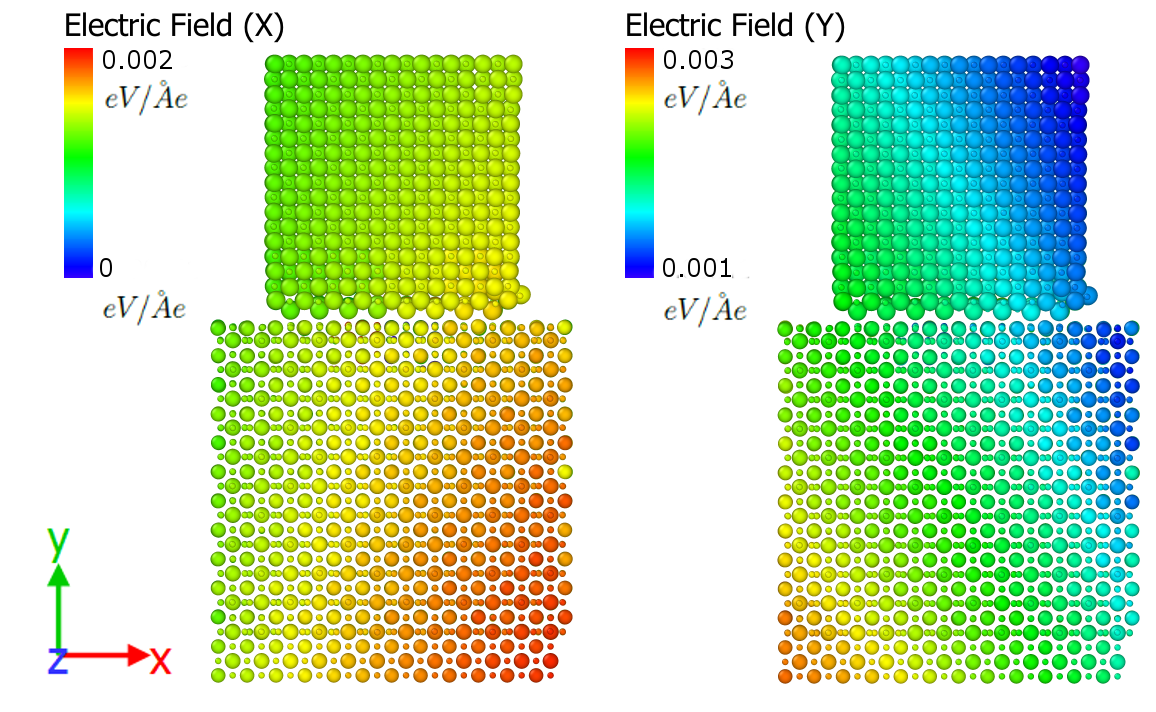}
\caption{Evolution of the distribution of the X and Y components of the electric field}\label{fig:electricfield}
\end{figure}

The potential difference between the vertical ends of the two slabs is shown in Figure \ref{fig:potentialdiffevolutiony}, which shows the average potential difference after final equilibration to be -4.4 mV. In the setting of a typical contact-separation TENG, these are the locations where a conductive electrode would be placed. The resistance of the air gap formed between the two materials during separation will prevent the charges from flowing between the slabs and the two conductive electrodes would be connected to a load so the operation of the TENG can power it \cite{Niu2015}. Figure \ref{fig:electricfield} shows the distribution of the X and Y components of the induced electric field, where the Z component distribution is similar to the X one. The presence of the electric field will induce charge transfer in the conductive electrodes as a direct result to contact electrification \cite{Wang2017b,Niu2015}. Also, The field distributions are shown to be consistent with the potential distribution in the sense that the field will exert a force on the charges to move from higher to lower potential zones.

\section{Conclusion}

The main cause of triboelectric charging in dielectrics is largely debatable, which complicates the determination of the direction of charge transfer between dielectrics in contact. Even for pre-specified material pairs, the direction of charge transfer can change because of differences in the nature of contact, temperature or microstructure among other factors. This work presented an AFT-based atomistic formulation for triboelectricity in dielectrics which relates the formation of the surface dipoles to the deformations of the surface lattices. The surface dipoles are theorized to be one cause for the triboelectric effect. First, the formulation is derived from AFT and basic principles of electrostatics. Thereafter, the formulation is used for the calculation of the electric characteristics (potential, field and charge density) by processing the output of an MD simulation case of a BaTiO$_3$/MgO tribopair. The results confirm the surface occurrence of the triboelectric effect as well as its relation to the contact-induced lattice deformations. It was also found based on the calculations that a BaTiO$_3$/MgO tribopair would be able to attain an electric potential difference of 104 $V/cm^2$ of the slabs which compared well with recently obtained experimental values found in the literature. Additionally, the electric field was presented to confirm the effect of the surface dipole formations on all electrical aspects of the system. Such high output could be the driving force for charge transfer in triboelectrification or contact electrification.

\section*{Acknowledgement}
This research was supported in part by the U.S. National Science Foundation, grant number 1662879. The authors gratefully acknowledge this support. 



\section*{Reference}
\bibliographystyle{model1-num-names}
\bibliography{references.bib}







\end{document}